\documentclass[12pt]{article}

\usepackage{amsmath}
\usepackage{bm}
\def\beq{\begin{equation}}
\def\eeq{\end{equation}}
\def\beqa{\begin{eqnarray}}
\def\eeqa{\end{eqnarray}}
\def\bea*{\begin{eqnarray*}}
\def\eea*{\end{eqnarray*}}
\def\bc{\begin{center}}
\def\ec{\end{center}}
\def\bar{\overline}

\def\a{\alpha}
\def\b{\beta}

\def\G{\varGamma}
\def\d{\delta}
\def\D{\varDelta}
\def\e{\epsilon}
\def\th{\theta}
\def\k{\kappa}
\def\l{\lambda}
\def\L{\varLambda}
\def\n{\nu}
\def\m{\mu}
\def\th{\theta}

\def\r{\rho}

\def\M{{\cal M}}
\def\N{{\cal N}}
\def\O{{\cal O}}
\def\Z{{\cal Z}}
\def\H{{\cal H}}
\bmdefine{\bw}{w}
\bmdefine{\bm}{m}
\bmdefine{\bz}{z}
\bmdefine{\bh}{h}
\bmdefine{\by}{y}

\def\PR#1#2#3{Phys. Rev.  {\bf #1}, #2 (#3)}
\def\PRL#1#2#3{Phys. Rev. Lett. {\bf #1}, #2 (#3)}
\def\PL#1#2#3{Phys. Lett. {\bf #1}, #2 (#3)}
\def\NP#1#2#3{Nucl. Phys. {\bf #1}, #2 (#3)}

\def\PTP#1#2#3{Prog. Theor. Phys. \textbf{#1}, #2 (#3)}

\begin{document}

\begin{center}
{\Large \bf Maki-Nakagawa-Sakata matrix \\
and Froggatt-Nielsen mechanism}

\vspace{15mm}

Chuichiro Hattori $^{\rm{1},}$
            \footnote{E-mail: hattori@aitech.ac.jp}, 
Mamoru Matsunaga $^{\rm{2},}$
            \footnote{E-mail: matsuna@phen.mie-u.ac.jp}, and
Takeo Matsuoka $^{\rm{2},}$
            \footnote{E-mail: t-matsu@siren.ocn.ne.jp}

\end{center}

\begin{center}
\textit{
$^{\rm{1}}$Science Division, General Education, Aichi Institute of Technology, \\
     Toyota 470-0392, Japan \\
$^{\rm{2}}$Department of Physics Engineering, Mie University, \\
     Tsu 514-8507, Japan
 }
\end{center}

\vspace{5mm}

\begin{abstract}
Flavor mixings together with fermion mass spectra are studied in detail in the 
$SU(6) \times SU(2)_R$ model, 
in which state-mixings beyond the minimal supersymmetric standard model (MSSM) take place. 
Characteristic patterns of fermion spectra are attributed to the hierarchical structure of 
effective Yukawa interactions via the Froggatt-Nielsen mechanism and also to 
the state-mixings beyond the MSSM. 
It is shown explicitly that the neutrino mass matrix in the mass diagonal basis 
for charged leptons has no hierarchical structure. 
This is due to the cancellation among the hierarchical factors by the seesaw mechanism 
with the hierarchical Majorana mass matrix of $R$-handed neutrinos.  
As a consequence, $V_{\rm MNS}$ exhibits large mixing. 
It is found that observed values of the Maki-Nakagawa-Sakata matrix elements are reproduced successfully. 
\end{abstract}

\vspace{10mm}


\section{INTRODUCTION}

Understanding the characteristic structure of fermion masses and flavor mixings is 
one of the major outstanding problems in particle physics. 
Up-type quarks, down-type quarks, charged leptons and neutrinos have distinct 
hierarchical mass patterns from each other. 
Moreover, the observed flavor mixing is small for quarks but large for leptons. 
This situation is in sharp contrast to a naive expectation from quark-lepton 
unification. 
We can disentangle ourselves from this discordance by considering the state-mixings 
between quarks (leptons) and extra particles beyond the minimal supersymmetric 
standard model (MSSM). 
In fact, it was shown in the context of $SU(6) \times SU(2)_R$ string-inspired model, 
which contains massless particles beyond the MSSM, that we are able to explain 
characteristic patterns of the observed mass spectra and mixing matrices of 
quarks and leptons \cite{Matsu1,Matsu2,Matsu3,Matsu4,Matsu5}. 
In the model the Froggatt-Nielsen (F-N) mechanism \cite{F-N} plays an important role. 
In Ref.\cite{Matsu5} it was predicted that the absolute value of the (1, 3) element 
of the Maki-Nakagawa-Sakata (MNS) matrix \cite{MNS} $|U_{e3}| = |\sin \th_{13}|$ lies between 
$\l$ and $\l^2$, where $\l = 0.23$. 
Recently, new data on neutrino mixings became available \cite{th13}. 
The observed value of the mixing angle $|\sin \th_{13}|$ turned out to be $\sim \l^{1.3}$, 
which supports this prediction.

In this paper we carry out more detailed study of fermion mass hierarchies 
and flavor mixings in the $SU(6) \times SU(2)_R$ string-inspired model. 
The hierarchical structure due to the F-N mechanism comes out 
not only in the effective Yukawa couplings but also in the $R$-handed neutrino 
Majorana mass matrix. 
In the neutrino sector, where the seesaw mechanism \cite{Seesaw} is at work, 
the hierarchical factors (the F-N factors) in the Dirac mass matrix are faced with 
the inverse of those in the Majorana mass matrix. 
This situation in the neutrino sector brings about a significant feature 
peculiar to the MNS matrix, on which the main emphasis is placed in this paper. 
As shown later, the neutrino mass matrix takes the form 
\beq
  \M_{\n} \propto \L_{\k} \, \varSigma \, \L_{\k} 
\label{neutrino}
\eeq
in the mass diagonal basis for charged leptons with 
\beqa
  \L_{\k} & = &  {\rm diag}\left( \k_1, \ \k_2, \ 1 \right), \\
  \varSigma & \simeq & \left(
                       \begin{array}{ccc}
                            1      &      0      &     0      \\
                       -\sigma_1   &      1      &     0      \\
                        \sigma_3   &  -\sigma_4  &     1              
                       \end{array}
                        \right)  \times N^{-1} \times 
                  \left(
                       \begin{array}{ccc}
                            1      &   -\sigma_1   &   \sigma_3    \\
                            0      &        1      &  -\sigma_4    \\
                            0      &        0      &     1              
                       \end{array}
                        \right). 
\label{Sgm}
\eeqa
Our prediction contains that both the parameters $\k_i \ (i = 1, \ 2)$ and 
$\sigma_i \ (i = 1, \ 3, \ 4)$ are $\O(1)$. 
The above form of $\M_{\n}$ is derived as a consequence of the fact that the F-N 
factors appearing in the charged lepton masses cancel out in large part through the seesaw 
mechanism. 
The matrix $N$ in Eq.(\ref{Sgm}), in which the F-N factors are eliminated from 
$R$-handed neutrino Majorana mass matrix, has no hierarchical structure and 
$\det N = 1$.

Taking simply $N = \L_{\k} = {\bf 1}$ and putting $\sigma_1 =2.30$, $\sigma_3 = 1.19$, 
and $\sigma_4 = 2.09$, we obtain the MNS matrix 
\[
  V_{\rm MNS} = \left(
                       \begin{array}{ccc}
                       \     0.875 \   &  \  0.453  \  &  \  0.170    \\
                       \    -0.431 \   &  \  0.570  \  &  \  0.699    \\
                       \     0.220 \   &  \ -0.685  \  &  \  0.695    
                       \end{array}
                        \right) 
\]
for the normal hierarchy. 
All elements of $V_{\rm MNS}$ offered here accommodate the present experimental 
data \cite{PDG} within 20\% provided that $\d_{CP} = 0$. 
In addition, eigenvalues of $\varSigma$ lead to 
\[
   \left( \frac{\D m_{32}^2}{\D m_{21}^2} \right)^{1/2} = 5.56, 
\]
which is in accord with the observed value. 
It is also found that the present framework is consistent with all observed values of 
fermion masses and mixings.

This paper is organized as follows. 
In Sec.~II we briefly review the $SU(6) \times SU(2)_R$ string-inspired model 
together with the F-N mechanism. 
Taking the mass matrix of up-type quarks as an example, we illustrate the whole scheme 
of the present model. 
For comparison with the case of the lepton sector, 
we study the mass matrix of down-type quarks in Sec.~III. 
State-mixings occur between down-type quarks and colored Higgsinos with even $R$-parity. 
We give the explicit form of the Cabibbo-Kobayashi-Maskawa (CKM) matrix, which proves to exhibit small mixing. 
Similarly to the case of down-type quarks, state-mixings take place between leptons and 
$SU(2)_L$-doublet Higgsinos. 
The mass matrix in the charged lepton sector is studied in Sec.~IV.\; 
Sec.~V deals with the neutrino sector in which state-mixings enter into the 
seesaw mechanism. 
The present approach provides a phenomenological framework, which enable us to 
analyze many experimental data. 
Numerical analysis of the MNS matrix and fermion spectra is given in Sec.~VI.\; 
Sec.~VII is devoted to summary.

\vspace{5mm}

\section{MODEL AND FROGGATT-NIELSEN MECHANISM}

In this study we choose $SU(6) \times SU(2)_R$ as a unification gauge 
symmetry at the underlying string scale $M_S$, which can be derived from 
the perturbative heterotic superstring theory via the flux breaking \cite{Matsu1}. 
In terms of $E_6$ we set matter superfields which consist of three families and 
one vector-like multiplet, i.e., 
\beq
  3 \times {\bf 27}(\Phi_{1,2,3}) + 
        ({\bf 27}(\Phi_0)+\overline{\bf 27}({\bar \Phi})) .
\eeq
The superfields $\Phi$ are decomposed into two multiplets of $G= SU(6) \times SU(2)_R$ as 
\beq
  \Phi({\bf 27})=\left\{
       \begin{array}{lll}
         \phi({\bf 15},{\bf 1}) & : & \quad \mbox{$\{Q,L,g,g^c,S\}$}, \\
         \psi(\overline{\bf 6},{\bf 2}) & : & \quad \mbox{$\{(U^c,D^c),(N^c,E^c),(H_u,H_d)\}$}, 
       \end{array}
       \right.
\eeq
where $g$, $g^c$ and $H_u$, $H_d$ represent colored Higgs and $SU(2)_L$-doublet 
Higgs superfields, respectively. 
Doublet Higgs and color-triplet Higgs fields belong to  different representations of 
$G$ and  this situation is favorable to solve the triplet-doublet splitting problem. 
The superfields $N^c$ and $S$ are $R$-handed neutrinos and $SO(10)$-singlets, respectively. 
Although $D^c$ and $g^c$ as well as $L$ and $H_d$ have the same quantum numbers under 
the standard model gauge group $G_{\rm SM} = SU(3)_c \times SU(2)_L \times U(1)_Y$, 
they belong to different irreducible representations of $G$. 
We assign odd $R$-parity to superfields $\Phi_{1,2,3}$ and even to $\Phi_0$ and 
$\bar{\Phi}$, respectively. 
Since ordinary Higgs doublets have even $R$-parity, they are contained in $\Phi_0$. 
It is assumed that $R$-parity remains unbroken down to the electroweak scale.

The gauge symmetry $G$ is spontaneously broken in two steps at the scales 
$\langle S_0\rangle=\langle \bar{S} \rangle$ and 
$\langle N_0^c\rangle=\langle \bar{N^c} \rangle$ as 
\beq
   G = SU(6) \times SU(2)_R 
     \buildrel \langle S_0 \rangle \over \longrightarrow 
             SU(4)_{\rm PS} \times SU(2)_L \times SU(2)_R  
     \buildrel \langle N^c_0 \rangle \over \longrightarrow 
     G_{\rm SM}, 
\eeq
where $SU(4)_{\rm PS}$ represents the Pati-Salam $SU(4)$ \cite{Pati}. 
The $D$-flatness conditions require $\langle S_0\rangle=\langle {\overline S} \rangle$ 
and $\langle N_0^c\rangle=\langle {\overline N^c} \rangle$ at each step of 
the symmetry breakings. 
Hereafter it is supposed that the symmetry breaking scales are 
$\langle S_0 \rangle = 10^{17 - 18}$GeV and $\langle N^c_0 \rangle = 10^{15 - 17}$GeV. 
Under the $SU(4)_{\rm PS} \times SU(2)_L \times SU(2)_R$ the chiral superfields 
$\phi({\bf 15},{\bf 1})$ and $\psi(\overline{\bf 6},{\bf 2})$ are decomposed as 
\begin{eqnarray*}
   ({\bf 15},{\bf 1})         &=& {\bf (4,2,1)} + {\bf (6,1,1)} + {\bf (1,1,1)}, \\
   (\overline{\bf 6},{\bf 2}) &=& {\bf (\overline{4},1,2)} + {\bf (1,2,2)}, 
\end{eqnarray*}
where each matter field is assigned as 
\begin{eqnarray*} 
\phi {\bf (4,2,1)}   &:& \ \ Q, L, \\
\phi {\bf (6,1,1)}   &:& \ \ g, g^c, \\
\phi {\bf (1,1,1)}   &:& \ \ S, \\
\psi {\bf (\overline{4},1,2)} &:& \ \ U^c, D^c, N^c, E^c, \\
\psi {\bf (1,2,2)}   &:& \ \ H_u, H_d. 
\end{eqnarray*} 
At the first step of the symmetry breaking the $R$-parity even fields $Q_0$, $L_0$, 
${\overline Q}$, ${\overline L}$ and $(S_0 - {\overline S})/\sqrt{2}$ 
are absorbed by gauge fields. 
Through the subsequent symmetry breaking the fields $U_0^c$, $E_0^c$, 
${\overline U}^c$, ${\overline E}^c$ and 
$(N_0^c - {\overline N}^c)/\sqrt{2}$ are absorbed.

{}From the viewpoint of the string unification theory, it is probable 
that the hierarchical structure of Yukawa couplings is attributed to 
some kind of flavor symmetries at the string scale $M_S$. 
If the flavor symmetry contains Abelian groups, 
the F-N mechanism works for the interactions among quarks, leptons and Higgs 
fields \cite{F-N}. 
The superpotential at the string scale is governed by the flavor symmetry as well 
as the gauge symmetry $G$. 
Aside from the flavor symmetry, we have two types of gauge invariant trilinear 
combinations 
\beqa
    (\phi ({\bf 15},{\bf 1}))^3 & = & QQg + Qg^cL + g^cgS, \nonumber \\
    \phi ({\bf 15},{\bf 1})(\psi (\overline{\bf 6},{\bf 2}))^2 & 
            = & QH_dD^c + QH_uU^c + LH_dE^c  + LH_uN^c \\
             {}& & \qquad   + SH_uH_d + gN^cD^c + gE^cU^c + g^cU^cD^c.  \nonumber 
\eeqa
They must be multiplied by additional $G$-invariant factors to form flavor 
symmetric terms in the superpotential at the string scale. 
For instance, the couplings associated with up-type quark fields are given by 
the nonrenormalizable terms 
\beq
    \widetilde{W}_U  = \sum_{i,j=1}^{3} \, m_{ij} 
        \left( \frac{S_0 \, \bar{S}}{M_S^2} \right)^{\mu_{ij}}  Q_i U^c_j H_{u0} 
\label{Wu}
\eeq
with dimensionless parameters $m_{ij} = \O(1)$, 
where the subscripts $i$ and $j$ are the generation indices and the exponent 
${\mu_{ij}}$'s are controlled by the flavor symmetry \cite{Matsu1,Matsu6,Matsu7}.

Here we add a few words about the flavor symmetries. 
It is known that all nongauge symmetries break down around the Planck scale 
due to quantum gravity effects \cite{Banks}. 
Therefore, it would be natural for the flavor symmetry to be unbroken discrete 
subgroups of local gauge symmetries. 
If this is the case, the discrete symmetry would be stable with respect to quantum 
gravity effects and then remains in the low-energy effective theory. 
Such a discrete flavor symmetry should be nonanomalous \cite{Ibanez}. 
For the present we denote the discrete flavor symmetry by $G_F$. 
Because the gauge symmetry at the string scale $M_S$ is assumed to be 
$SU(6) \times SU(2)_R$, 
the mixed anomaly conditions $G_F \cdot (SU(6))^2$ and $G_F \cdot (SU(2)_R)^2$ 
are imposed on the $G_F$-charges of the massless matter fields. 
The heavy fermions decouple in these anomalies but not in the cubic $G_F^{\ 3}$ and 
the mixed $G_F \cdot ({\rm Graviton})^2$ anomalies. 
However, we have no information about the heavy modes. 
Hence the cubic $G_F^{\ 3}$ and the mixed $G_F \cdot ({\rm Graviton})^2$ anomaly 
conditions are not relevant to the constraints on the flavor charges of matter 
fields in the low-energy effective theory. 
Anomaly-free solutions in the $SU(6) \times SU(2)_R$ model have been explored 
previously in Ref. \cite{Matsu7}. 
In this paper we concentrate our attention on fermion mixings and their 
spectra.

At low energies effective Yukawa interactions arise from $G$-invariant 
nonrenormalizable terms which respect the flavor symmetry. 
For example, Eq.(\ref{Wu}) leads to the effective Yukawa interactions 
\beq
    W_U  = \sum_{i,j=1}^{3} \, \M_{ij} Q_i U^c_j H_{u0} 
\eeq
for up-type quarks with 
\beq
   \M_{ij} = m_{ij} 
     \left( \frac{\langle S_0 \rangle \, \langle \bar{S} \rangle }{M_S^2} \right)^{\mu_{ij}}. 
\eeq
Due to the F-N mechanism, the dimensionless matrix $\M$ takes the form 
\beq
             \M = f_M \, \G_1 M \G_2. 
\label{Mnoh}
\eeq
Our basic assumption is that the hierarchical structure of all $3 \times 3$ mass matrices 
is attributed to the F-N factors $\G_1$ and/or $\G_2$. 
Hence, hierarchy of $\M$ stems only from $\G_1$ and $\G_2$, and the dimensionless 
matrix $M$ contains no hierarchical structure. 
Here we put a factor $f_M$ in order to set $\det M = 1$. 
The F-N factors $\G_1$ and $\G_2$ are described as 
\beq
    \G_1 = {\rm diag}( x^{\a_1}, \ x^{\a_2}, \ 1), \qquad 
             \G_2 = {\rm diag}( x^{\b_1}, \ x^{\b_2}, \ 1), 
\eeq
where $x$ is given by 
\[
   x = \frac{\langle S_0 \rangle \, \langle \bar{S} \rangle}{M_S^2} < 1 
\]
and $(S_0{\bar S})$ is a $G$-invariant with a nonzero flavor charge. 
The exponents $\a_1, \ \a_2, \ \b_1, \ \b_2$ in the F-N factors are determined 
by assigning flavor charges to the matter fields. 
Even if $x$ in itself is not a very small number, physical parameters can be 
very small if they depend on high powers of $x$. 
We assume the hierarchical patterns 
\beq
     x^{\a_1} \ll x^{\a_2} \ll 1, \qquad x^{\b_1} \ll x^{\b_2} \ll 1 
\eeq
by suitably chosen flavor charges. 
In this paper we ignore the phase factors of vacuum expectation values (VEV's).

The mass matrix $\M$ is diagonalized via biunitary transformation as 
\beq
    {\cal V}_u^{-1} \M \,{\cal U}_u = \L_u, \qquad 
                   v_{u0} \L_u = {\rm diag}(m_u, \ m_c, \ m_t) 
\eeq
with $v_{u0} = \langle H_{u0} \rangle$. 
Up-type quark masses are given by 
\beq
   (m_u, \ m_c, \ m_t) \, \simeq \, v_{u0} \, f_M \times 
             \left( \frac{1}{\bar{m}_{11}} \, x^{\a_1 + \b_1}, \ 
                     \frac{\bar{m}_{11}}{m_{33}} \, x^{\a_2 + \b_2}, \  m_{33} \right), 
\label{umass}
\eeq
where $m_{ij} = (M)_{ij}$ and $\bar{m}_{ij} = \D(M)_{ij}^*$ \cite{Matsu1,Matsu2,Matsu3}. 
$\D(M)_{ij}$'s are the cofactors of $M$. 
The diagonalization matrix is described in terms of column vectors $\bw_i^{(u)}$ 
$(i = 1,2,3)$ as 
\beq
   {\cal V}_u = ( \bw_1^{(u)}, \ \bw_2^{(u)}, \ \bw_3^{(u)}), 
\eeq
where $\bw_i^{(u)}$'s are eigenvectors of $\M \M^{\dag}$ and expressed as  
\[
      \bw_1^{(u)} = N_1^{(u)} \left(
              \begin{array}{c}
                     1      \\
                 u_1^{(u)}  \\
                 v_1^{(u)}  
              \end{array}
              \right),  \quad 
     \bw_2^{(u)} = N_2^{(u)} \left(
              \begin{array}{c}
                 u_2^{(u)}  \\
                     1      \\
                 v_2^{(u)}  
              \end{array}
              \right),  \quad 
     \bw_3^{(u)} = N_3^{(u)} \left(
              \begin{array}{c}
                 u_3^{(u)}  \\
                 v_3^{(u)}  \\
                     1      
              \end{array}
              \right) 
\]
with 
\[
\begin{array}{ll}
   u_1^{(u)} \simeq x^{\a_1 - \a_2} \ \displaystyle\frac{\bar{m}_{21}}{\bar{m}_{11}}, 
                      \qquad \qquad 
     &  v_1^{(u)} \simeq x^{\a_1} \ \displaystyle\frac{\bar{m}_{31}}{\bar{m}_{11}},   \\[4mm]
   u_2^{(u)} \simeq - (u_1^{(u)})^*, \qquad \qquad 
     &  v_2^{(u)} \simeq - x^{\a_2} \ \displaystyle\frac{m_{23}^*}{m_{33}^*},     \\[4mm]
   u_3^{(u)} \simeq x^{\a_1} \ \displaystyle\frac{m_{13}}{m_{33}}, \qquad \qquad 
     &  v_3^{(u)} \simeq  - (v_2^{(u)})^* 
\end{array}
\]
and 
\[
   \qquad N_i^{(u)} = \left( 1 + \left| u_i^{(u)} \right|^2 
                        + \left| v_i^{(u)} \right|^2 \right)^{-1/2}, \quad (i = 1, \ 2, \ 3). 
\]

\vspace{5mm}


\section{THE CKM MATRIX}

For comparison with the case of the lepton sector, we study the mass matrix of 
down-type quarks in this section. 
At energies below the scale $\langle N_0^c \rangle$ there appear mixings 
between $D^c$ and $g^c$ which are $SU(2)_L$-singlets.\footnotemark[1] 
Effective Yukawa interactions among down-type colored fields are of the form 
\beq
    W_D = \sum_{i,j=1}^{3} \, \left[ \Z_{ij} \ g_i g^c_j S_0 + 
          \M_{ij} \left( g_i D^c_j N^c_0 + Q_i D^c_j H_{d0} \right) \right], 
\eeq
where the dimensionless matrix ${\cal Z} = f_Z \, \G_1 Z \G_1$ and $\det Z = 1$. 
As explained in the previous section, there is no hierarchical structure in $Z$. 
The mass matrix of down-type colored fields is given by the $6 \times 6$ 
matrix \cite{Matsu1,Matsu2,Matsu3} 
\beq
   \begin{array}{r@{}l} 
       \vphantom{\bigg(}   &  \begin{array}{ccc} 
            \quad  \ g^c   &  \quad \  D^c  &  
   \end{array}  \\ 
   \widehat{\M}_d = 
      \begin{array}{l} 
        g   \\  D  \\ 
      \end{array} 
       & 
   \left( 
      \begin{array}{cc} 
          \r_S \Z    &    \r_N \M   \\
             0       &    \r_d \M   
      \end{array} 
\right) 
\end{array} 
\label{Md}
\eeq
in unit of $M_S$, where $\r_S = \langle S_0 \rangle /M_S$, 
$\r_N = \langle N^c_0 \rangle /M_S$ and 
$\r_d = \langle H_{d0} \rangle /M_S = v_{d0} /M_S$. 
\footnotetext[1]{An early attempt of explaining the CKM matrix via $D^c$--$g^c$ mixings 
has been made in Ref.\cite{Hisano}, in which a SUSY $SO(10)$ model was taken.}

The above mass matrix $\widehat{\M}_d$ is diagonalized via biunitary 
transformation as 
\beq
    \widehat{\cal V}_d^{-1} \widehat{\M}_d \, \widehat{\cal U}_d = 
                   {\rm diag}(\L_d^{(0)}, \ \e_d \, \L_d^{(2)}), 
\eeq
where $\e_d = \r_d / \r_N = v_{d0} / \langle N^c_0 \rangle = \O(10^{-15})$. 
To solve the eigenvalue problem, it is convenient to take 
$\widehat{\M}_d \widehat{\M}_d^\dag$ and express it as 
\beq
   \widehat{\M}_d \widehat{\M}_d^{\dag} =  \left(
  \begin{array}{cc}
     A_d + B_d   &   \e_d B_d   \\
     \e_d B_d    &  \e_d^2 B_d  
  \end{array}
  \right),
\eeq
where $A_d = |\r_S|^2 \, \Z \Z^{\dag}$ and $B_d = |\r_N|^2 \, \M \M^{\dag}$.
Since $\e_d$ is a very small number, we can carry out our calculation 
by using perturbative $\e_d$-expansion. 
Among six eigenvalues three of them represent heavy modes with the grand unified theory (GUT) scale masses. 
The remaining three, corresponding to down-type quarks, are derived from diagonalization 
of the $\e_d^2 \, (A_d^{-1} + B_d^{-1})^{-1}$, namely 
\beq
   \e_d^2 \, {\cal V}_d^{-1} (A_d^{-1} + B_d^{-1})^{-1} {\cal V}_d = 
                      (\e_d \, \L_d^{(2)})^2. 
\eeq
Down-type quark masses are given by 
$M_S \, \e_d \, \L_d^{(2)} = {\rm diag}(m_d, \ m_s, \ m_b)$. 
Explicit forms are 
\beq
   (m_d, \ m_s, \ m_b) \, \simeq \, v_{d0} \, f_M \, x^{\b_1} \times 
          \left( \frac{1}{\sqrt{a}} \, x^{\a_1}, \ 
               \sqrt{\frac{a}{b}} \, x^{\a_2}, \ 
                   \sqrt{\frac{b}{c}} \right), 
\label{dmass}
\eeq
where 
\beqa
    a & = & \xi_d^2 \, |\bar{z}_{11}|^2 + |\bar{m}_{11}|^2,   \qquad \qquad 
    b \ = \ \xi_d^2 \, |(\bar{\bz}_1 \times \bar{\bm}_1)_3|^2,   \nonumber   \\[2mm]
\label{qadc}
    c & = & \xi_d^4 \, x^{2(\a_1 - \a_2)} \, |({\bz}_3^* \cdot \bar{\bm}_1)|^2 \, + \, 
            \xi_d^2 \, x^{2(\b_1 - \b_2)} \, |({\bm}_3^* \cdot \bar{\bz}_1)|^2,  \\[-1mm]
    &  &  \xi_d^2 \ = \ \left| \frac{\r_N f_M}{\r_S f_Z} \right|^2 \, x^{2(\b_1 - \a_1)}.  \nonumber 
\eeqa
Here we use the notations $z_{ij} = (Z)_{ij}$, $\bar{z}_{ij} = \D(Z)_{ij}^*$ and 
\beqa
    &  &  {\bm}_i = ( m_{1i}, \ m_{2i}, \ m_{3i})^T, \qquad \ 
              \bar{\bm}_i = ( \bar{m}_{1i}, \ \bar{m}_{2i}, \ \bar{m}_{3i})^T, 
                                                                        \nonumber  \\
    &  &  {\bz}_i = ( z_{1i}, \ z_{2i}, \ z_{3i})^T, \qquad \qquad \ 
              \bar{\bz}_i = ( \bar{z}_{1i}, \ \bar{z}_{2i}, \ \bar{z}_{3i})^T.   \nonumber  
\eeqa
The diagonalization matrix ${\cal V}_d$ is of the form 
\beq
   {\cal V}_d = ( \bw_1^{(d)}, \ \bw_2^{(d)}, \ \bw_3^{(d)}), 
\eeq
\[
     \bw_1^{(d)} = N_1^{(d)} \left(
              \begin{array}{c}
                     1      \\
                 u_1^{(d)}  \\
                 v_1^{(d)}  
              \end{array}
              \right),  \quad 
     \bw_2^{(d)} = N_2^{(d)} \left(
              \begin{array}{c}
                 u_2^{(d)}  \\
                     1      \\
                 v_2^{(d)}  
              \end{array}
              \right),  \quad 
     \bw_3^{(d)} = N_3^{(d)} \left(
              \begin{array}{c}
                 u_3^{(d)}  \\
                 v_3^{(d)}  \\
                     1      
              \end{array}
              \right), 
\]
where 
\[
\begin{array}{ll}
   u_1^{(d)} \simeq x^{\a_1 - \a_2} \ \displaystyle\frac
             {\xi_d^2 \, \bar{z}_{21}\,\bar{z}_{11}^* + \bar{m}_{21}\,\bar{m}_{11}^*}
                   {\xi_d^2 \, |\bar{z}_{11}|^2 + |\bar{m}_{11}|^2}, \qquad 
     & v_1^{(d)} \simeq x^{\a_1} \ \displaystyle\frac
             {\xi_d^2 \, \bar{z}_{31}\,\bar{z}_{11}^* + \bar{m}_{31}\,\bar{m}_{11}^*}
                   {\xi_d^2 \, |\bar{z}_{11}|^2 + |\bar{m}_{11}|^2},  \\[4mm]
   u_2^{(d)} \simeq - (u_1^{(d)})^*,  
     & v_2^{(d)} \simeq - x^{\a_2} \ \displaystyle\frac{(\bar{\bz}_1 \times \bar{\bm}_1)_2}
                                     {(\bar{\bz}_1 \times \bar{\bm}_1)_3}, \\[4mm]
   u_3^{(d)} \simeq x^{\a_1} \ \displaystyle\frac{(\bar{\bz}_1 \times \bar{\bm}_1)_1^*}
                   {(\bar{\bz}_1 \times \bar{\bm}_1)_3^*}, 
     & v_3^{(d)} \simeq - (v_2^{(d)})^* 
\end{array}
\]
and 
\[
  \qquad N_i^{(d)} = \left( 1 + \left| u_i^{(d)} \right|^2 
                   + \left| v_i^{(d)} \right|^2 \right)^{-1/2}, \quad (i = 1, \ 2, \ 3). 
\]

We are now in a position to calculate the CKM mixing matrix \cite{CKM} as 
\beq
    V_{\rm CKM} = {\cal V}_u^{-1} \,{\cal V}_d = {\cal V}_u^{\dag} \,{\cal V}_d. 
\label{VCKM}
\eeq
Thus 
\beq
    (V_{\rm CKM})_{ij} = {\bw}_i^{(u)*} \cdot {\bw}_j^{(d)}. 
\eeq
More explicitly, we have \cite{Matsu3} 
\beqa
  V_{us} = (V_{\rm CKM})_{12}  & \simeq &  x^{\a_1 - \a_2} \ 
      \frac{\xi_d^2 \, \bar{z}_{11} \, (\bar{\bz}_1 \times \bar{\bm}_1)_3^*}
           {\bar{m}_{11}^* \, a},  \nonumber    \\[-4mm]
  V_{cb} = (V_{\rm CKM})_{23}  & \simeq &  x^{\a_2} \ 
      \frac{\bar{m}_{11}^* \, ({\bm}_3 \cdot \bar{\bz}_1^*)}
           {m_{33} \, (\bar{\bz}_1 \times \bar{\bm}_1)_3^* }, \nonumber    \\[-2mm]
\label{ckm}
  V_{cd} = (V_{\rm CKM})_{21}  & \simeq &  - (V_{us})^*, \\[2mm]
  V_{ts} = (V_{\rm CKM})_{32}  & \simeq &  - (V_{cb})^*,  \nonumber  \\[2mm]
  V_{td} = (V_{\rm CKM})_{31}  & \simeq &  (V_{us} \, V_{cb})^*,  \nonumber  \\[-2mm]
  V_{ub} = (V_{\rm CKM})_{13}  & \simeq &  x^{3\a_1 - 2\a_2} \ 
      \frac{\xi_d^2 \, z_{33}^* \, ({\bz}_3 \cdot \bar{\bm}_1^*)}
        {\bar{m}_{11}^* \, \left| (\bar{\bz}_1 \times \bar{\bm}_1)_3 \right|^2}. 
                                                                  \nonumber 
\eeqa
All off-diagonal elements of the $V_{\rm CKM}$ contain the F-N factors. 
This means that there is little difference between the diagonalization matrices 
for up-type quarks and for down-type quarks in $SU(2)_L$-doublets. 
Further, $V_{ub}$ is zero in the leading order but nonzero in 
the next-to-leading order, which implies that the element $V_{ub}$ is 
naturally suppressed compared to $V_{td}$.

We now take up the case that each first term in Eq.(\ref{qadc}) for $a$ and $c$ is dominant. 
This case is realized for the parameter choice preferred in numerical analysis later. 
From Eqs.(\ref{umass}), (\ref{dmass}) and (\ref{ckm}), we have 
\beq
   \frac{m_s}{m_u} \, V_{us} \simeq \frac{v_{d0}}{v_{u0}}, 
\eeq
which is independent of values of $\a_i$ and $\b_i$ $(i = 1, \ 2)$. 
The observed value of the left-hand side leads to $(v_{d0}/v_{u0}) \simeq 10.5$, 
which is in marked contrast to the usual solution with large $\tan \b = v_u/v_d$. 
This fact suggests that there exist Higgs fields other than $H_{u0}$ and $H_{d0}$ 
and that some or all of them develop their VEV's.

\vspace{5mm}


\section{CHARGED LEPTON MASS MATRIX}

Effective Yukawa interactions among charged leptons are described as 
\beq
   W_E = \sum_{i,j=1}^{3} \, \left[ \H_{ij} H_{di} H_{uj} S_0 + 
       \M_{ij} \left( L_i H_{uj} N^c_0 + L_i E^c_j H_{d0} \right) \right], 
\eeq
where the dimensionless matrix $\H = f_H \, \G_2 H \G_2$ with $\det H = 1$. 
The matrix $H$ has no hierarchical structure. 
In the lepton sector, mixings occur between $L$ and $H_d$ which are $SU(2)_L$-doublets. 
Consequently, the charged lepton mass matrix is expressed in terms of 
the $6 \times 6$ matrix \cite{Matsu4,Matsu5} 
\beq
    \begin{array}{r@{}l} 
       \vphantom{\bigg(}   &  \begin{array}{ccc} 
           \quad \ H_u^+   &  \quad  E^{c+}  &  
    \end{array}  \\ 
    \widehat{\M}_l = 
       \begin{array}{l} 
            H_d^-  \\  L^-  \\ 
       \end{array} 
       & 
    \left( 
      \begin{array}{cc} 
          \r_S \, \H   &        0      \\
           \r_N \M     &     \r_d \M   
      \end{array} 
    \right) 
   \end{array} 
\eeq 
in unit of $M_S$. 
The study of the charged lepton mass matrix is parallel to that of 
the down-type quark mass matrix in the previous section. 
The matrix $\widehat{\M}_l$ is diagonalized via biunitary transformation as 
\beq
    \widehat{\cal V}_l^{-1} \widehat{\M}_l \, \widehat{\cal U}_l = 
                   {\rm diag}(\L_l^{(0)}, \ \e_d \, \L_l^{(2)}). 
\eeq
Among six eigenvalues three of them represent heavy modes with the GUT scale masses. 
The remaining three, corresponding to charged leptons, are derived from the diagonalization 
\beq
   \e_d^2 \, {\cal V}_l^{-1} (A_l^{-1} + B_l^{-1})^{-1} {\cal V}_l = 
                      (\e_d \, \L_l^{(2)})^2, 
\eeq
where $A_l = |\r_S|^2 \, \H^{\dag} \H$ and $B_l = |\r_N|^2 \, \M^{\dag} \M$. 
Charged lepton masses are given by 
$M_S \, \e_d \, \L_l^{(2)} = {\rm diag}(m_e, \ m_{\m}, \ m_{\tau})$. 
Explicit forms are 
\beq
   (m_e, \ m_{\m}, \ m_{\tau}) \, \simeq \, v_{d0} \, f_M \, x^{\a_1} \times 
          \left( \frac{1}{\sqrt{a'}} \, x^{\b_1}, \ 
               \sqrt{\frac{a'}{b'}} \, x^{\b_2}, \ 
                   \sqrt{\frac{b'}{c'}} \right), 
\label{chlma}
\eeq
where 
\beqa
    a' & = & \xi_l^2 \, |\bar{h'}_{11}|^2 + |\bar{m'}_{11}|^2,   \qquad \qquad 
    b' \ = \ \xi_l^2 \, |(\bar{\bh'}_1 \times \bar{\bm'}_1)_3|^2,  \nonumber \\[2mm]
\label{ladc}
    c' & = & \xi_l^4 \, x^{2(\b_1 - \b_2)} \, 
                    |({\bh}_3^{'*} \cdot \bar{\bm'}_1)|^2 \, + \, 
                 \xi_l^2 \, x^{2(\a_1 - \a_2)} \, 
                    |({\bm}_3^{'*} \cdot \bar{\bh'}_1)|^2,  \\[-1mm]
    &  &  \xi_l^2 \ = \ \left| \frac{\r_N f_M}{\r_S f_H} \right|^2 \, x^{2(\a_1 - \b_1)} 
              \, = \, \xi_d^2 \, \left| \frac{f_Z}{f_H} \right|^2 \, x^{4(\a_1 - \b_1)}. \nonumber
\eeqa
Here we use the notations $\bar{m'}_{ij} = \D(M)_{ji} = \bar{m}_{ji}^*$, 
$h'_{ij} = (H^{\dag})_{ij}$, $\bar{h'}_{ij} = \D(H)_{ji}$ and 
\beqa
    &  &  {\bm}'_i = ( m_{i1}, \ m_{i2}, \ m_{i3})^{\dag}, \qquad \ 
              \bar{\bm'}_i = ( \bar{m}_{i1}, \ \bar{m}_{i2}, \ \bar{m}_{i3})^{\dag}, 
                                                                        \nonumber  \\[2mm]
    &  &  {\bh}'_i = ( h'_{1i}, \ h'_{2i}, \ h'_{3i})^T, \qquad \qquad \ 
              \bar{\bh'}_i = ( \bar{h'}_{1i}, \ \bar{h'}_{2i}, \bar{h'}_{3i})^T.   \nonumber  
\eeqa
Hereafter we introduce the notation 
\beq
  \L_{\k} = {\rm diag}\left( \sqrt{\frac{c'}{a' \,b'}}, \ \ 
                   \frac{\sqrt{c' \, a'}}{b'}, \ \ 1 \right) 
                     = {\rm diag}\left( \k_1, \ \k_2, \ 1 \right), 
\label{Lk}
\eeq
where the parameters $\k_i\;(i = 1, \ 2)$ are $\O(1)$. 
Then Eq.(\ref{chlma}) is rewritten as 
\beq
   \L_l^{(2)} \simeq \r_N \, f_M \ x^{\a_1} \, \sqrt{\frac{b'}{c'}} \times \G_2 \times \L_{\k}. 
\label{chlmb}
\eeq

The diagonalization matrix is of the form 
\beq
   {\cal V}_l = ( {\bw}_1^{(l)}, \ {\bw}_2^{(l)}, \ {\bw}_3^{(l)}), 
\label{Vchal}
\eeq
\[
      {\bw}_1^{(l)} = N_1^{(l)} \left(
              \begin{array}{c}
                     1      \\
                 u_1^{(l)}  \\
                 v_1^{(l)}  
              \end{array}
              \right),  \quad 
     {\bw}_2^{(l)} = N_2^{(l)} \left(
              \begin{array}{c}
                 u_2^{(l)}  \\
                     1      \\
                 v_2^{(l)}  
              \end{array}
              \right),  \quad 
     {\bw}_3^{(l)} = N_3^{(l)} \left(
              \begin{array}{c}
                 u_3^{(l)}  \\
                 v_3^{(l)}  \\
                     1      
              \end{array}
              \right). 
\]
Each element is given by 
\[
\begin{array}{ll}
  u_1^{(l)}   \simeq   \sigma_1 \ x^{\b_1 - \b_2},  \qquad \qquad \ 
    & v_1^{(l)}   \simeq   \sigma_2 \ x^{\b_1},      \\[3mm]
  u_2^{(l)}   \simeq  - (u_1^{(l)})^*,  
    & v_2^{(l)}   \simeq   \sigma_4 \ x^{\b_2},       \\[3mm]
  u_3^{(l)}   \simeq   \sigma_3^* \ x^{\b_1},  
    & v_3^{(l)}   \simeq  - (v_2^{(l)})^*  
\end{array}
\]
and 
\[
  \qquad  N_i^{(l)} = \left( 1 + \left| u_i^{(l)} \right|^2 
                           + \left| v_i^{(l)} \right|^2 \right)^{-1/2} \quad (i = 1, \ 2, \ 3), 
\]
where 
\beq
\begin{array}{ll}
   \sigma_1 =  \displaystyle\frac
         {\xi_l^2 \, \bar{h'}_{21}\,\bar{h'}_{11}^* + \bar{m'}_{21}\,\bar{m'}_{11}^*}
              {\xi_l^2 \, |\bar{h'}_{11}|^2 + |\bar{m'}_{11}|^2},  \qquad & 
   \sigma_2 = \displaystyle\frac
         {\xi_l^2 \, \bar{h'}_{31}\,\bar{h'}_{11}^* + \bar{m'}_{31}\,\bar{m'}_{11}^*}
              {\xi_l^2 \, |\bar{h'}_{11}|^2 + |\bar{m'}_{11}|^2},  \\[8mm]
   \sigma_3 =  \displaystyle\frac{(\bar{\bh'}_1 \times \bar{\bm'}_1)_1}
                   {(\bar{\bh'}_1 \times \bar{\bm'}_1)_3}, & 
   \sigma_4 = - \displaystyle\frac{(\bar{\bh'}_1 \times \bar{\bm'}_1)_2}
                                     {(\bar{\bh'}_1 \times \bar{\bm'}_1)_3}. 
\end{array}
\eeq
Note that the parameters $\sigma_i\;(i = 1 \-- 4)$ are $\O(1)$.

\vspace{5mm}


\section{NEUTRINO MASS MATRIX}

In the neutral lepton sector we have the $15 \times 15$ mass matrix \cite{Matsu4,Matsu5} 
\beq
\begin{array}{r@{}l} 
   \vphantom{\bigg(}   &  \begin{array}{cccccc} 
          \quad \, H_u^0   & \quad \  H_d^0  &  \quad \ L^0  
                                 &  \quad \ \ N^c   &  \quad \  S  &
        \end{array}  \\ 
\widehat{{\cal M}}_{NS} = 
   \begin{array}{l} 
        H_u^0  \\  H_d^0  \\  L^0  \\  N^c  \\  S  \\
   \end{array} 
     & 
\left( 
  \begin{array}{ccccc} 
       0      & \r_S \, \H  &   \r_N \M^T   &      0     &  \r_d \M^T    \\
  \r_S \, \H  &      0      &       0       &      0     &  \r_u \M^T    \\
   \r_N \M    &      0      &       0       &  \r_u \M   &       0       \\
       0      &      0      &  \r_u \M^T    &     \N     &       0       \\
   \r_d \M    &  \r_u \M    &       0       &      0     &   {\cal S}    \\
  \end{array} 
\right), 
\end{array} 
\label{NM}
\eeq
where $\r_u = \langle H_{u0} \rangle /M_S = v_{u0} /M_S$ and 
$\N, \ {\cal S}$ stand for Majorana mass matrices for the superfields $N^c$ and $S$ 
with odd $R$-parity. 
This mass matrix comes from the effective Yukawa interactions 
\beqa
   W_{NS} & = & \sum_{i,j=1}^{3} \, \H_{ij} H_{di} H_{uj} S_0 + 
           \ \sum_{i,j=1}^{3} \, \M_{ij} 
                    \left( L_i H_{uj} N^c_0 + L_i N^c_j H_{u0} \right) \nonumber \\[-5mm]
             & & \qquad \qquad  + \ \sum_{i,j=1}^{3} \, \M_{ij} 
                     \left( S_i H_{uj} H_{d0} + S_i H_{dj} H_{u0} \right) 
\eeqa
and from Majorana mass terms for $N^c$ and $S$. 
Here we assume (the electroweak scale) $\ll$ ($\N$ scale) $\ll$ (${\cal S}$ scale). 
The matrix $\N$ is symmetric and has the form 
\[
   \N = f_N \, \G_2 N \G_2, 
\]
in which $N$ contains no hierarchical structure and $\det N = 1$.

Mixings in the lepton sector occur between the $SU(2)_L$-doublet fields $L$ and $H_d$. 
When we diagonalize the charged lepton mass matrix, the neutral leptons in $SU(2)_L$-doublet 
undergo the same transformation as the diagonalization matrix for charged leptons. 
In addition, the seesaw mechanism is at work. 
Hence, the neutrino mass matrix for light modes becomes 
\beq
  \M_{\n} = M_S \, \e_u^2 \, \L_l^{(2)} \, {\cal V}_l^{\dag} \, 
                                   \N^{-1} \, {\cal V}_l^* \, \L_l^{(2)} 
\label{Mn}
\eeq
in the diagonal mass basis for charged leptons, 
where $\e_u = \r_u / \r_N = v_{u0} / \langle N^c_0 \rangle = \O(10^{-15})$. 
By diagonalizing $\M_{\n}$ we obtain neutrino masses 
\beq
   {\cal V}_{\n}^T \, \M_{\n} \, {\cal V}_{\n} = 
                        {\rm diag}(m_{\n_1}, \ m_{\n_2}, \ m_{\n_3}) 
\eeq
and the MNS matrix \cite{MNS} 
\beq
    V_{\rm MNS} = {\cal V}_{\n}^T. 
\eeq
The matrix $\M_{\n}$ is rewritten as 
\beq
    \M_{\n} = \frac{M_S \, \e_u^2}{f_N} \ Y^T \, N^{-1} \, Y, 
\eeq
where 
\beq
    Y = \G_2^{-1} \, {\cal V}_l^* \, \L_l^{(2)}. 
\eeq
As seen in Eq.(\ref{Vchal}), ${\cal V}_l$ is of the form 
\beq
  {\cal V}_l \simeq \left(
     \begin{array}{ccc}
                  1                 &  - \sigma_1^* \,x^{\b_1 - \b_2}  &    \sigma_3^* \,x^{\b_1}  \\
        \sigma_1 \,x^{\b_1 - \b_2}  &               1                  &  - \sigma_4^* \,x^{\b_2}  \\
            \sigma_2 \,x^{\b_1}     &       \sigma_4 \,x^{\b_2}        &           1               
     \end{array}
                        \right). 
\label{Vl}
\eeq
In addition, the matrix $\L_l^{(2)}$ is given by Eq.(\ref{chlmb}). 
It is important to remember that the parameters $\sigma_i\;(i = 1 \-- 4)$ and 
$\k_i\;(i = 1, \ 2)$ in ${\cal V}_l$ and $\L_{\k}$ are $\O(1)$. 
Hence, $Y$ is described as 
\beq
  Y \simeq \r_N \, f_M \ x^{\a_1} \, \sqrt{\frac{b'}{c'}} \times 
               (\G_2^{-1} \, {\cal V}_l^* \, \G_2) \times \L_{\k} 
\label{Y}
\eeq
and 
\beq
  \G_2^{-1} \, {\cal V}_l^* \, \G_2 = 
              \left(
                \begin{array}{ccc}
                            1                    &         -\sigma_1        &   \sigma_3    \\
                \sigma_1^* \,x^{2(\b_1 - \b_2)}  &             1            &  -\sigma_4    \\
                    \sigma_2^* \,x^{2\b_1}       &  \sigma_4^* \,x^{2\b_2}  &       1       
                \end{array}
              \right). 
\label{GVG}
\eeq
It is noticeable that in the upper triangular elements of the above matrix the F-N factors 
$x^{\b_1}$ and $x^{\b_2}$ cancel out. 
This cancellation mechanism comes from the feature that both $(N^c, \ E^c)$ and 
$(H_u, \ H_d)$ belong to $(\overline{\bf 6},{\bf 2})$ representation of 
$G= SU(6) \times SU(2)_R$. 
It follows that 
\beq
  \M_{\n} = \frac{v_{u0}^2 \, f_M^2}{M_N} \, 
                  x^{2 \a_1} \, \frac{b'}{c'} \times \L_{\k} \, \varSigma \, \L_{\k}, 
\label{MNe}
\eeq
where 
\beq
   \varSigma = (\G_2^{-1} \, {\cal V}_l^* \, \G_2)^T \, N^{-1} \, 
                            (\G_2^{-1} \, {\cal V}_l^* \, \G_2). 
\label{Sigma}
\eeq
$M_N = f_N \, M_S$ represents the typical scale of the $R$-handed neutrino Majorana masses. 
In the present model the F-N factors $x^{2 \b_1}, \ x^{2 \b_2}$ and $x^{2 (\b_1-\b_2)}$ 
are sufficiently small compared to 1. 
Therefore, $\varSigma$ is approximated as 
\beq
  \varSigma \simeq \left(
                       \begin{array}{ccc}
                            1      &      0      &     0      \\
                       -\sigma_1   &      1      &     0      \\
                        \sigma_3   &  -\sigma_4  &     1              
                       \end{array}
                        \right)  \times N^{-1} \times 
                  \left(
                       \begin{array}{ccc}
                            1      &   -\sigma_1   &   \sigma_3    \\
                            0      &        1      &  -\sigma_4    \\
                            0      &        0      &     1              
                       \end{array}
                        \right). 
\label{Pi}
\eeq
Incidentally, Cholesky decomposition says that a general positive definite symmetric 
matrix can be decomposed as $L L^T$ uniquely by using an appropriate lower triangular 
matrix $L$ with positive diagonal elements. 
Since the matrix $N$ is symmetric, it is possible to implement the Cholesky decomposition 
provided that $N$ is positive definite. 
Consequently, redefining the parameters $\sigma_i$ $(i = 1, \ 3, \ 4)$ and 
$\k_i$ $(i = 1, \ 2)$, we can put $N = {\bf 1}$ without loss of generality. 
In the following analysis we take $N = {\bf 1}$. 
As stated in Sec.~IV, our prediction contains that both Eq.(\ref{Pi}) and $\L_{\k}$ 
are described in terms of $\O(1)$ parameters. 
The MNS matrix $V_{\rm MNS}$ is the diagonalization matrix of $\L_{\k} \varSigma \L_{\k}$. 
Accordingly, the texture of matrices $H$ and $N$, which are defined by removing 
the F-N factors, governs neutrino mass ratios and the MNS matrix.

\vspace{5mm}


\section{NUMERICAL ANALYSIS}

Experimental data show that the neutrino mass ratio is \cite{PDG} 
\beq
   r = \left( \frac{\D m_{32}^2}{\D m_{21}^2} \right)^{1/2} = 5.56 \pm 0.20. 
\eeq
Hereafter, in the analysis we settle $r = 5.56$ and search realistic solutions of 
the MNS matrix for the normal hierarchy, requiring $2m_1 \leq m_2 \leq \frac{1}{2}m_3$. 
Let us first consider a simple case $\L_{\k} = {\bf 1}$. 
In this case the matrix $V_{\rm MNS}$ is nothing but the diagonalization matrix of $\varSigma$, 
in which there are three parameters $\sigma_i$ $(i = 1, \ 3, \ 4)$. 
Under the condition $r = 5.56$ there remain only two independent parameters. 
Realistic solutions consonant with the sign texture of the MNS matrix are obtained 
in the parameter region $\sigma_1, \ \sigma_3, \ \sigma_4 > 0$. 
For example, when $\sigma_1 =2.30$, $\sigma_3 = 1.19$, and $\sigma_4 = 2.09$, 
the MNS matrix is estimated as 
\beq
  V_{\rm MNS} = \left(
                       \begin{array}{ccc}
                       \     0.875 \   &  \  0.453  \  &  \  0.170    \\
                       \    -0.431 \   &  \  0.570  \  &  \  0.699    \\
                       \     0.220 \   &  \ -0.685  \  &  \  0.695    
                       \end{array}
                        \right).
\eeq
Experimental data provided that $\d_{CP} = 0$ show \cite{PDG} 
\beq
  V_{\rm MNS}^{({\rm PDG})} = \left(
                       \begin{array}{ccc}
                       \    0.824 \   & \   0.547 \   &  \ 0.145    \\
                       \   -0.499 \   & \   0.583 \   &  \ 0.641    \\
                       \    0.266 \   & \  -0.601 \   &  \ 0.754    
                       \end{array}
                        \right) 
\eeq
for the normal hierarchy. 
Although we take a very simple case $\L_{\k} = {\bf 1}$, 
all elements offered here are consistent with the observed values within 20\% 
including $|(V_{\rm MNS})_{13}| = |\sin \th_{13}|$. 
The ratio of the neutrino masses turn out to be $ 0.019 : 1.00 : 5.64$. 
The observed absolute value of $\D m_{32}^2 \simeq m_{\n_3}^2$ is obtained 
by taking $M_N = \O(10^9){\rm GeV}$. 
Instead of the simple choice $\L_{\k} = {\bf 1}$, we next consider the case 
$\L_{\k} = {\rm diag}\left( 1, \ \k_2, \ 1 \right)$. 
In this case we have three parameters. 
When we put $\k_2 = 0.639$ together with $\sigma_1 = 2.80$, $\sigma_3 = 1.23$, 
and $\sigma_4 = 1.72$, 
all elements of the calculated MNS matrix are in agreement with the observed values 
within 1\%. 
The mass ratio of neutrinos becomes $ 0.021 : 1.00 : 5.65$.

Finally we discuss charged fermion mass spectra and the CKM matrix. 
The setting $\L_{\k} = {\bf 1} \ (\k_1 = \k_2 = 1)$ in Eq.(\ref{chlmb}) means 
\beq
   m_e : m_{\m} : m_{\tau} = x^{\b_1} : x^{\b_2} : 1. 
\eeq
From experimental values of charged lepton masses we have 
$x^{\b_1} = \l^{5.6}$ and $x^{\b_2} = \l^{1.9}$ with $\l = 0.23$. 
As to the other parameters, we set 
$x^{\a_1} = \l^{3.3}$,  $x^{\a_2} = \l^{2.3}$ for the F-N factor $\G_1$ 
and $\xi_d = \l^{-0.2}$, $\xi_l = \l^{-5}$ which leads to 
\beq
    f_Z = \frac{\r_N}{\r_S} \, f_M \, \l^{2.5}, \qquad 
    f_H = \frac{\r_N}{\r_S} \, f_M \, \l^{2.7},
\eeq
respectively. 
In this setting each first term in Eq.(\ref{qadc}) for $a$ and $c$, and in Eq.(\ref{ladc}) 
for $a'$ and $c'$ is dominant. 
As pointed out in Sec.~III, it follows that $(v_{d0}/v_{u0}) \simeq 10.5 = \l^{-1.6}$. 
In the present setting most of fermion mass hierarchies are attributed to the F-N factors. 
In order to reproduce observed mass spectra and the CKM matrix precisely, 
we adjust the other parameters as 
\beqa
\label{param}
 &  &  m_{33} = \l^{0.7}, \qquad z_{33} = \l^{-0.3}, \nonumber \\
 &  &  \bar{m}_{11} = \l^{0.5}, 
                    \qquad \bar{z}_{11} = \l^{-0.3}, \qquad \bar{h'}_{11} = \l^{2.9}, \nonumber \\
 &  &  |(\bar{\bz}_1 \times \bar{\bm}_1)_3| = \l^{0.2}, \qquad 
          |({\bz}_3^* \cdot \bar{\bm}_1)| = \l^{0.2}, \qquad 
               |({\bm}_3 \cdot \bar{\bz}_1^*)| = \l^{0.3},   \\ 
 &  &  |(\bar{\bh'}_1 \times \bar{\bm'}_1)_3| = \l^{0.8}, \qquad 
          |({\bh'}_3^* \cdot \bar{\bm'}_1)| = \l^{0.0}. \nonumber 
\eeqa
As seen in Tables I and II, our results are in good agreement with observed values. 
Although there are many parameters, 
it is noteworthy that these parameters other than $\bar{h'}_{11}$ are $\O(1)$. 
Note that this parameter $\bar{h'}_{11}$ stands for the cofactor $\D(H)_{11}$.

\vspace{10mm}

\begin{table}[ht]
\bc
\caption{Quark and lepton masses divided by top quark mass ($\l = 0.23$)}

\vspace{2mm}

\begin{tabular}{c c c} \hline\hline
\vphantom{\LARGE I}   Mass ratio   &  \quad Our result \quad  &  Observed values    \\ \hline
\vphantom{\LARGE I}  $m_c/m_t$       &    $\l^{3.3}$   &    $\l^{3.34}$      \\
\vphantom{\LARGE I}  $m_u/m_t$       &    $\l^{7.7}$   &    $\l^{7.65}$      \\ \hline
\vphantom{\LARGE I}  $m_b/m_t$       &    $\l^{2.5}$   &    $\l^{2.54}$      \\
\vphantom{\LARGE I}  $m_s/m_t$       &    $\l^{5.1}$   &    $\l^{5.11}$      \\
\vphantom{\LARGE I}  $m_d/m_t$       &    $\l^{7.1}$   &    $\l^{7.14}$      \\ \hline
\vphantom{\LARGE I}  $m_{\tau}/m_t$  &    $\l^{3.1}$   &    $\l^{3.12}$      \\
\vphantom{\LARGE I}  $m_{\m}/m_t$    &    $\l^{5.0}$   &    $\l^{5.04}$      \\
\vphantom{\LARGE I}  $m_e/m_t$       &    $\l^{8.7}$   &    $\l^{8.67}$      \\ \hline\hline
\end{tabular}
\ec
\end{table}

\begin{table}[ht]
\bc
\caption{Elements of the CKM matrix}

\vspace{2mm}

\begin{tabular}{c c c} \hline\hline
\vphantom{\LARGE I}  $(V_{\rm CKM})_{ij}$  &  \quad Our result \quad  &  Observed values    \\ \hline
\vphantom{\LARGE I}   $V_{us}$          &    $\l^{1.0}$   &    $\l^{1.02}$      \\
\vphantom{\LARGE I}   $V_{cb}$          &    $\l^{2.2}$   &    $\l^{2.17}$      \\ 
\vphantom{\LARGE I}   $V_{td}$          &    $\l^{3.2}$   &    $\l^{3.23}$      \\
\vphantom{\LARGE I}   $V_{ub}$          &    $\l^{3.9}$   &    $\l^{3.87}$      \\ \hline\hline
\end{tabular}
\ec
\end{table}

\vspace{5mm}


\section{SUMMARY AND DISCUSSION}

We have studied flavor mixings, especially the MNS matrix, in detail 
in the $SU(6) \times SU(2)_R$ model, in which the F-N mechanism plays an important role. 
Due to the F-N mechanism both effective Yukawa couplings and $R$-handed neutrino Majorana 
mass matrix have hierarchical structure, which is described in terms of the F-N factors. 
In this model the $D^c$--$g^c$ and $L$--$H_d$ mixings as well as generation mixings occur 
and affect both fermion mass spectra and flavor mixings.

In the $D^c$--$g^c$ mixings, since $D^c$ and $g^c$ are both $SU(2)_L$-singlets, 
the disparity between the diagonalization matrices for up-type quarks and 
down-type quarks in $SU(2)_L$-doublets is rather small. 
Accordingly, $V_{\rm CKM}$ exhibits small mixing. 
On the other hand, in the $L$--$H_d$ mixings, since $L$ and $H_d$ are both 
$SU(2)_L$-doublets, there appears no disparity between the diagonalization matrices 
for charged leptons and neutrinos unless the seesaw mechanism does not take place. 
As a matter of fact, however, the seesaw mechanism is at work and an additional 
transformation is required to diagonalize the neutrino mass matrix. 
This additional transformation matrix yields nontrivial $V_{\rm MNS}$. 
In the present model the neutrino mass matrix has characteristic structure 
as seen in Eqs.(\ref{MNe}) and (\ref{Pi}). 
A noticeable point derived from the present model is the fact that both $\varSigma$ 
and $\L_{\k}$ are described in terms of $\O(1)$ parameters. 
This is due to the mechanism that the F-N factors cancel out in the upper triangular 
elements of $Y$. 
As a consequence, there is no hierarchical structure in $\M_{\n}$ and eventually 
$V_{\rm MNS}$ exhibits large mixing. 
We have shown that observed values of MNS matrix elements can be reproduced successfully. 
Neutrino mass ratios and the MNS matrix are subject directly to the texture of the matrices 
$H$ and $N$ which are free from the F-N factors. 
It is expected that the study of neutrino mass ratios and the MNS matrix provides 
an important clue as to what symmetry governs the matrices $H$ and $N$.

The characteristic structure of fermion spectra is attributed to the hierarchical 
effective Yukawa couplings due to the F-N mechanism and also to the $D^c$--$g^c$ 
and $L$--$H_d$ mixings. 
In particular, the difference of mass hierarchies among up-type quarks, down-type quarks 
and charged leptons has its origin in $D^c$--$g^c$ and $L$--$H_d$ mixings. 
In the neutrino sector we have to incorporate the Majorana mass hierarchy of 
$R$-handed neutrinos with the seesaw mechanism. 
Numerical results are consistent with all observed values of fermion masses 
and mixings. 
The present model provides a unified description of mass spectra and flavor mixings.

It is worth making reference to three interesting relations among fermion masses 
and the CKM mixing matrix in which the F-N factors cancel out. 
Two of them are 
\beqa
   \frac{m_b^2 \, V_{cb} V_{ub}}{m_s^2 \, V_{us}}     & = & \O(1), 
\label{bs}                                                         \\
   \frac{m_d \, m_b \, V_{ub}}{m_s^2 \, (V_{us})^3}   & = & \O(1) 
\label{dbs}
\eeqa
for the quark sector. 
Further, in the third relation lepton masses are related to quark masses 
and $V_{cb}$ as 
\beq
   \frac{m_e \, m_{\mu}}{m_{\tau}} \, \frac{m_t}{ m_u \, m_s} \, V_{cb}  = \O(1). 
\label{etau}
\eeq
Although fermion masses and the CKM matrix elements contain the F-N factors, 
the above products in Eqs.(\ref{bs}), (\ref{dbs}) and (\ref{etau}) are free 
from the F-N factors. 
Namely, provided that all parameters in Eq.(\ref{param}) other than $\bar{h'}_{11}$ 
are $\l^{0.0}$, 
the left-hand side of the above equations become unity. 
The observed values in Eqs.(\ref{bs}) and (\ref{dbs}) are 1.19 and 0.67, respectively. 
These relations are consistent with the experimental data. 
It is remarkable that the observed values in Eq.(\ref{etau}) are nearly equal to 1.00. 
In other words, this implies an additional relation among $\O(1)$ parameters, 
which is expressed as 
\beq
  \frac{ |({\bm}_3 \cdot \bar{\bz}_1^*)|}{|\bar{z}_{11}|} = 
     \frac{|(\bar{\bh'}_1 \times \bar{\bm'}_1)_3|^2}
         {|\bar{m}_{11}|^2 \, |({\bh'}_3^* \cdot \bar{\bm'}_1)|}. 
\eeq
The interrelations among the matrices $M$, $Z$, and $H$ may lie behind this relation.

In Sec.~III we have pointed out that in the present framework $(v_{d0}/v_{u0})$ is large. 
This result contrasts sharply with the usual solution with large $\tan \b = v_u/v_d$ 
and suggests that Higgs fields other than $H_{u0}$ and $H_{d0}$ develop their VEV's. 
It is expected that there exist rich spectra of Higgs fields beyond those of the MSSM 
at the TeV scale.

In order to determine the F-N factors and the magnitude of $f_i$'s $(i = M, \ Z, \ H, \ N)$ 
theoretically, we need an appropriate flavor symmetry and also the flavor charge assignment 
to matter fields. 
As stated in Sec.~II, many of the $R$-parity even chiral superfierlds are absorbed by 
gauge superfields. 
Remaining $R$-parity even chiral superfields are colored Higgs fields ($g_0$, $g^c_0$, 
$\overline{g}$, $\overline{g^c}$), $SU(2)_L$-doublet Higgs fields ($H_{u0}$, $H_{d0}$, 
$\overline{H}_u$, $\overline{H}_d$), and $G_{\rm SM}$-singlet fields 
($(S_0 + \overline{S})/\sqrt{2}$, $(N^c_0 + \overline{N^c})/\sqrt{2}$. 
The colored Higgs fields possibly get their masses at the GUT scale under an appropriate 
flavor symmetry. 
Consequently, the $R$-parity even chiral superfields which are available at the TeV scale 
are only $SU(2)_L$-doublet Higgs fields and $G_{\rm SM}$-singlet fields. 
Detailed spectra of matter fields depend on the flavor symmetry and also 
the flavor charge assignment to matter fields. 
In the previous works we made an attempt to find several solutions \cite{Matsu1, Matsu6, Matsu7}. 
The detailed study of this issue is the subject of future works. 
Further, in this paper we ignored the phase factors of VEV's for matter fields and 
assumed $\d_{CP} = 0$. 
The study of the CP-violation will be carried out elsewhere.

\vspace{5mm}


\end{document}